\documentclass[conference,10pt]{IEEEtran}

\usepackage{times,graphicx,subfigure, amsmath}
\usepackage{epstopdf,comment,color,cite,url}
\usepackage{algorithmic}
\usepackage{algorithm}
\definecolor{Red}{RGB}{255,0,0} 
\definecolor{Blue}{RGB}{0,0,255} 
\definecolor{Green}{RGB}{0,255,0} 

\newcommand{\Nc}{{\cal N}}

\newcommand{\nuh}{\hat{\nu}}
\newcommand{\etah}{\hat{\eta}}

\newcommand{\ddsG}{{{}_{G}}}

\begin{document}

\hyphenation{throu-gh-put}
\hyphenation{off-load}
\hyphenation{re-que-st}
\hyphenation{re-spon-se}
\hyphenation{al-lo-ca-tion}
\hyphenation{han-do-ver}
\hyphenation{com-mand}
\hyphenation{med-ium}
\hyphenation{so-ur-ces}
\hyphenation{whi-ch}
\hyphenation{al-lo-ca-tion-re-que-st}

\pagestyle{plain}

\title{An Energy Efficient Protocol for Gateway-Centric\\ Federated Residential Access Networks} 

\author{Claudio Rossi, Claudio Casetti, Carla-Fabiana Chiasserini\\
Politecnico di Torino, Italy
}

\maketitle 
 
\begin{abstract}
The proliferation of overlapping, always-on IEEE 802.11 Access Points (APs) in urban areas
can cause spectrum sharing conflicts, inefficient bandwidth usage and power waste.
Cooperation among APs could address these problems (i) by allowing
under-used devices to hand over their clients to nearby APs and
temporarily switch off, (ii) by balancing the load of clients among
APs and thus offloading congested APs.
The federated houses model provides an appealing backdrop to implement cooperation
among APs. In this paper, we outline a framework that, assuming the presence of
a multipurpose gateway with AP capabilities in every household, allows such cooperation
through the monitoring of local wireless resources and the triggering of offloading requests toward
other federated gateways. We  then present simulation results in realistic settings that 
provide some insight on the capabilities of our framework. 
\end{abstract} 
 
\section{Introduction}


The growing popularity of appliances and consumer devices embedding a
WiFi interface has led to the proliferation of Access Points (APs) in public areas and
private homes alike. In the latter case, however, the deployment usually occurs
in an uncoordinated fashion, leading to overlapping coverage and spectrum
conflicts. Additionally, APs in private homes are usually underloaded and 
are left on around the clock, both a power waste and an unnecessary increase in
electromagnetic pollution. 

Federated homes, i.e., neighborhoods where network resources 
are shared and networked devices belonging to different users 
cooperate,  have the potential to 
solve the above problems by incorporating APs in smart Gateways
that handle all inward and outward network traffic. Gateways are
advanced home devices capable of offering wireless Internet access, storage, and
multimedia services including audio and video real-time streaming. 

In order to optimize the usage of the wireless medium, neighboring,  
federated Gateways with overlapping coverages  
should identify and optimally relocate the Wireless Stations (WSs) among themselves, 
and, possibly, turn themselves off if a subset of Gateways can 
adequately support the current load requested by the WSs. Also, an underloaded 
(or temporarily switched off) Gateway should be called upon for help by Gateways that 
experience a congested wireless medium, and  
associate some of their WSs. 

Such operations require that Gateways have self-load 
assessment capabilities and run inter-Gateway 
procedures for WS relocation.
Load estimation techniques can be classified as passive or active. The latter ones require to inject probing packets into the network and  estimate the traffic load based on the delay experienced by such packets. Probing packets, however, yield additional overhead, and could have a negative impact on data flows, especially in case of real-time traffic \cite{back}. 
We will therefore focus on passive techniques, which aim at estimating the traffic load by observing some meaningful metrics.
However, 
existing passive estimation techniques
are not mature to fully support multi-rate WLANs with
variable traffic patterns.  Metrics based either on the number of
associated WSs~\cite{sta}, the channel busy (or, equivalently, idle)
time~\cite{busy1,busy2}, or the aggregated BSS throughput~\cite{thru}, are
affected by the payload size and the data rate of the transmitted
packets.  It follows that such metrics may indicate the availability
of bandwidth when the saturation throughput has been already reached,
or, conversely, they may detect saturation in presence of available
bandwidth.

Other techniques, e.g.,~\cite{beacon}, either apply only to
self estimation of the downlink bandwidth availability or require
changes in the WSs.

As for solutions enabling Gateways to switch themselves off, 
centralized schemes have been proposed in \cite{green1,green2}. 
These solutions, however, are suitable for coverages resulting from controlled placement of the
Gateways, as is the case of big enterprises and college campuses, but
they are hardly fitting for a residential scenario where each Gateway 
is independently placed within a household. Other solutions to 
overcome capacity limitations of single APs have suggested the use of
TDMA techniques to let WSs access multiple APs at a
time~\cite{aggregation}, requiring, however,  modification in 
the WSs.

In this paper, we address the above issues by defining a
solution that applies to a multirate network and to
 generic traffic  scenarios. In particular, we introduce: (i) a metric
and a procedure that allow the Gateways a self-evaluation of their
load status; (ii) a metric and a procedure that let a Gateway gauge the 
impact of the association of one or more WSs relocated from  a
neighboring Gateway;
(iii) a distributed protocol for inter-Gateway communication  and WS relocation that refrains from 
non-standard operations at the WSs, as well as non-standard signalling between 
Gateway and WSs.


\section{Preliminaries\label{sec:preliminaries}}
{\bf System scenario.}   
We consider  $M$ residential units (e.g., houses or
apartments), each of them  equipped with a Gateway ($G_1,\ldots,G_M$) that
offers wireless Internet access through the 802.11 technology.
Adjacent Gateways use different channel frequencies and
each Gateway is equipped with two radio interfaces: one for 
communicating with the WSs in 
the BSS controlled by the Gateway, the other for listening to
different frequency channels whenever needed. 

The Gateways are federated, i.e., they can communicate and coordinate
with each other using an out-of-band channel, which is their backhaul Internet connection.
Note that we do not assume the presence of any central network
controller that manages WSs association to the Gateways.


The WSs that operate within the generic BSS can be sources or destinations of elastic or inelastic
traffic flows, i.e, flows that use either TCP
or UDP at the transport layer. At the MAC layer, the Gateway and the WSs  may transmit frames with
different payload size and their data rate may vary according to the
experienced channel propagation conditions.

Depending on the traffic load and on the number of associated WSs 
within the BSS they control, Gateways 
are said to be in Light, Heavy or Regular status. The  Light status corresponds
to an underloaded BSS: if its WSs could be relocated to other
BSSs, the Gateway could switch itself off and save energy. The Heavy status,
instead, characterizes an overloaded BSS, where some WSs should
associate to other BSSs so as to let the users receive the desired
throughput. A Gateway in Regular status neither can switch itself off nor
does it need to give some of its WSs away, while it might accommodate relocated WSs
within its BSS. 
In order to let the Gateways assess their status, we  assume they
carry out traffic measurements as described below.

{\bf Assumptions.} 
A Gateway can access the ``protocol type'' field in the IP
packets, and collect  statistics on elastic and inelastic traffic 
within its BSS.
The Gateway carries out such measurements over time intervals,
named {\em cycles}. A cycle is defined as the minimum between a time 
$T_{max}$ and the period  
needed to let (1) each active WS  successfully send at least one data frame carrying inelastic traffic, and (2) the Gateway successfully transmit at least one data frame carrying inelastic traffic to every WS for which it has data to send. 
The Gateway considers a WS to be active in  cycle $j$ if it successfully receives from the WS at least one data frame within the time $T_{max}$ since the current cycle starting time. Likewise, the Gateway is active in cycle $j$  if it has sent at least one frame within the cycle. 
In the following, we denote by 
$C(j)$ the duration of cycle $j$, by
$\Nc(j)$ the set of nodes (WSs and Gateway) that were active in the
cycle, and by $N(j)$ the cardinality of $\Nc(j)$. 

Then, like the mechanism
we described in \cite{wons11}, 
at each cycle $j$ and for each active WS $k$, the Gateway computes a
running average of the  uplink throughput 
for elastic and inelastic traffic of $k$, denoted by $\eta_k(j)$ and
$\nu_k(j)$, respectively.
Likewise, the Gateway computes a running average of its own downlink throughput for
both elastic and inelastic traffic, denoted by $\eta_\ddsG(j)$ and $\nu_\ddsG(j)$, respectively.

In addition, for each frame successfully transmitted by a WS or by the
Gateway itself, the Gateway observes the payload size for elastic/inelastic traffic and the used
data rates, and it computes the corresponding running averages:
$P^{(e)}_k(j)$, $P^{(i)}_k(j)$, and $R_k(j)$\footnote{For the
data rate, the Gateway stores only one value because automatic rate 
adaptation algorithms do not distinguish between elastic and inelastic
flows.} ($k \in \Nc(j)$). 
We will refer to all the above measurements the Gateway performs for a
WS as the WS's traffic profile.
Furthermore, the Gateway computes the running average  
of the data rate, $R(j)$, and of the 
payload size, $P(j)$, over all data frames, carrying either
elastic or inelastic traffic, that it successfully sends or receives.

We then introduce a fundamental quantity for 
our bandwidth monitoring algorithm. 
Let us consider cycle $j$. At the end of the cycle, the Gateway  computes the (aggregate) saturation throughput $S(j)$, as defined in~\cite{chatz}, which extends the original Bianchi's model~\cite{bianchi} in presence of errors due to channel propagation conditions:
\begin{equation}
S(j)=\frac{N(j) \tau(j) [1-\tau(j)]^{N(j)-1} P(j) (1-p_e(j))}{E[T(j)]}.
\label{eq:S}
\end{equation}
In (\ref{eq:S}), $\tau(j)$ is the probability that a node (either a WS
or the Gateway) accesses the medium at a generic time slot in cycle $j$,
$p_e(j)$ is the filtered average packet error rate, 
and $E[T(j)]$ is the average duration of a time interval in which an event occurs (namely, an empty slot, a successful transmission,  a transmission failed due to channel errors, or a collision). 
The expressions of $\tau(j)$ and $E[T(j)]$ can be derived following~\cite{chatz} 
and are reported in the Appendix for completeness, while $p_e(j)$ can be estimated by the Gateway
based on the modulations used for the transmissions in the $j$-th cycle, their associated signal-to-noise ratio, and assuming independent bit errors on the channel. 
Using (\ref{eq:S}), the Gateway computes  the average per-node throughput
under saturation conditions, as $S_n(j)=S(j)/N(j)$. Note that $S_n(j)$
represents the saturation throughput for a node with average behavior,
i.e., a node using a payload size $P(j)$ and a data rate $R(j)$.


\section{Bandwidth monitoring}
\label{sec:algo}

Here, we first present the algorithm that lets a Gateway assess its load status.
Then, we describe how a Gateway can reliably evaluate the impact on
its BSS of associating additional stations that other Gateways would
like to relocate.  Finally, we present simulation results showing the
effectiveness of our bandwidth monitoring approach.

\subsection{Gateway status assessment}

Consider a generic Gateway that at the end of the current cycle, say
$j$, wants to gauge the traffic load within the BSS it controls.
To do so, it follows Alg.~\ref{algo:ACE_Mon}. 
\begin{algorithm}[h!]
\caption{Gateway status assessment}
\label{algo:ACE_Mon}
\begin{algorithmic}[1]
\REQUIRE $\Nc(j)$, $S(j)$, $S_n(j)$, $\eta_k(j)$, $\nu_k(j)$
\ENSURE Gateway status
\medskip
\STATE $B(j) \leftarrow S(j)$
\FOR {$k\in \Nc(j)$ }
        \STATE $B(j) \leftarrow B(j)- \min \{\nu_k(j)+\eta_k(j), S_n(j)\}$
        \STATE $B(j) \leftarrow B(j)-
        \max\bigg\{0,[\nu_k(j)-S_n(j)]\frac{R(j)}{R_k(j)}\bigg\}$
\ENDFOR
	\IF{$B(j)/S(j) > T_L \AND (N(j)-1) < N_L$} 
                        \STATE Gateway in Light status   
        \ELSIF{$B(j)/S(j) < T_R$} 
                        \STATE Gateway in Heavy status  
        \ELSE 
                        \STATE Gateway in Regular status
        \ENDIF 
\end{algorithmic}
\end{algorithm}

The idea at the basis of the algorithm is that, due to the
per-packet fairness provided by the 802.11a/b/g distributed access
scheme, any node $k \in \Nc(j)$, such that $\eta_k(j)+ \nu_k(j) \leq
S_n(j)$, can 
transmit all its uplink traffic, both elastic and inelastic (line 3),
while the others  reach $S_n(j)$ and then share the remaining
bandwidth, if any (line 4).  As $S_n(j)$ refers to the average node behavior,
we weigh the bandwidth in excess of $S_n(j)$ used  by node
$k$ with $R(j)/R_k(j)$, thus accounting for the actual node data
rate (line 4). 
We also stress that, for each node, only inelastic traffic exceeding the saturation share is
considered; elastic traffic above saturation is instead neglected, 
since  it can adapt to bandwidth availability. 

At the end of the procedure, we compare the bandwidth 
available for inelastic traffic normalized to the
saturation throughput, $B(j)/S(j)$, against two
different thresholds, as follows. 
We consider the Gateway to be in Light status if $B(j)/S(j)>T_L$ and
the number of WSs 
associated to it is smaller than $N_L$,
and in Heavy status if $B(j)/S(j)<T_R$. The Gateway is in Regular status otherwise.

\subsection{b-metric computation}

Next, we want a Gateway to
assess if it can associate one or more stations that
other Gateways are trying to relocate, without harming the existing WSs.
To do so, a Gateway computes the bandwidth available for inelastic
traffic within its BSS, as if the relocated WSs were actually associated;
we name such a quantity b-metric. Again, we focus on inelastic traffic only.
For simplicity, the b-metric computation will be outlined
in the case where a single WS has to be relocated. The extension to
the case of multiple WSs is straightforward.

Let $G_m$ be the Gateway that evaluates the  bandwidth available for
inelastic traffic within its BSS, $j$ identifies the last cycle and $x$ 
is the WS that  another Gateway tries to relocate.
Through signaling exchange between  Gateways, $G_m$ may acquire the 
uplink throughput of $x$ for inelastic and elastic traffic, as well as
the downlink throughput that $x$ would like to receive. If this is not
possible,
the Gateway takes a conservative approach and assigns to the WS a
traffic demand equal to the value of saturation throughput.
Also, $G_m$ updates the set  $\Nc(j)$ by adding $x$.

In order to evaluate the throughput that $x$ would achieve and
its impact on the performance of inelastic flows involving other nodes, we have to estimate the
throughput that each active 
node can obtain with respect to the value it has experienced in cycle $j$. 
To do so, we adopt the procedure reported in Alg.~\ref{algo:ACE}.

\begin{algorithm}[h!]
\caption{b-metric evaluation}
\label{algo:ACE}
\begin{algorithmic}[1]
\REQUIRE $\Nc(j)$, $S(j)$, $S_n(j)$, $R(j)$, $\eta_k(j)$, $\nu_k(j)$,
$P^{(e)}_k(j)$, $P^{(i)}_k(j)$, $R_k(j)$ 
\ENSURE $\mbox{b}(j)$ 
\medskip
\STATE $\beta \leftarrow S(j)$
\FOR {$k\in \Nc(j)$ }
        \STATE $\beta  \leftarrow \beta - \min \{\nu_k(j)+\eta_k(j),
        S_n(j) \}$
	\STATE $\nuh_k(j) \leftarrow \min\{\nu_k(j),S_n(j)\}$
        \STATE $\etah_k(j) \leftarrow \min\{\eta_k(j),S_n(j)-\nuh_k(j)\}$
\ENDFOR
\STATE $\Nc_o$ $\leftarrow$ Sort$(k \in \Nc(j) \,
| \,\nu_k(j)+\eta_k(j)>S_n(j)$, $R_k(j))$ 
\STATE $\mbox{b}(j)=\beta$ 
\WHILE{$\beta>0$  and $\Nc_o \neq \emptyset$}
	\FOR {any  $k\in \Nc_o$ and $\beta>0$}
		\IF{$\nuh_k(j) < \nu_k(j)$}
                        \STATE $\delta  \leftarrow \min\bigg\{ \frac{P^{(i)}_k(j) R(j)}{C(j) R_k(j)}, \beta \bigg\}$
			\STATE $\nuh_k(j) \leftarrow \nuh_k(j) + \delta$
                        \STATE $\beta \leftarrow \beta - \delta$
                        \STATE $\mbox{b}(j) \leftarrow \mbox{b}(j) - \delta$
		\ELSIF{$\etah_k(j) < \eta_k(j)$}
                        \STATE $\delta  \leftarrow \min\bigg\{ \frac{P^{(e)}_k(j) R(j)}{C(j) R_k(j)}, \beta  \bigg\}$
                	\STATE $ \etah_k(j) \leftarrow \etah_k(j) + \delta$
                        \STATE $\beta \leftarrow \beta - \delta$
                \ELSE
                       \STATE $\Nc_o \leftarrow   \Nc_o\setminus k$ 
		\ENDIF
	\ENDFOR
\ENDWHILE
\IF {$\mbox{b}(j)/S(j)>T_A$} 
      \STATE  association of $x$ is possible
\ELSE 
      \STATE  association of $x$ is rejected
\ENDIF     
\end{algorithmic}
\end{algorithm}

According to the proposed algorithm,
the Gateway first computes the remaining bandwidth $\beta$ as the difference
between  the saturation throughput $S(j)$ and the sum of the shares consumed by  
the active nodes (line 3). Again,  due to the per-packet fairness provided by the
access scheme, each node share is given by the minimum between $S_n(j)$ and its total (elastic and inelastic) throughput, as measured by the Gateway in cycle $j$.  
Then, lines 4--5  report the amount of inelastic and elastic node throughput that can be
accommodated within the $S_n(j)$ share. 

We identify the set of nodes $\Nc_o$
whose total (elastic and inelastic)  throughput exceeds $S_n(j)$ (line 7). 
Considering one of these nodes at a time, we 
assume that it will get a fraction of the remaining bandwidth so as to transmit one 
additional packet of average size. While doing this, the node will
give priority to inelastic traffic. This occurs while (i) $\beta>0$ and (ii) there is at least one node for which the throughput experienced in cycle $j$ has not been reached yet (lines 9--24).
As $S_n(j)$ has been computed considering the average node behavior,
we weigh the bandwidth consumed by node $k$ to transmit a packet by $R(j)/R_k(j)$, thus accounting for the actual data rate used by the node (lines 12 and 17). 
Also, we consider the worst case in which nodes with the lowest data rate $R_k(j)$ seize the channel first. Indeed, the lower the data rate, the larger the consumed bandwidth (line 7).

The b-metric, $\mbox{b}(j)$, is initialized to $\beta$ (line 8) and
decreased by the estimated inelastic share of each active node that
exceeds $S_n(j)$ (line 15).  It thus corresponds to the bandwidth that is
still available for inelastic traffic within the BSS. Finally, the
association of WS $x$ is considered as possible only if $\mbox{b}(j)/S(j)>T_A$, where $T_A$ is a given threshold.
Note that,  a Gateway always accepts 
association requests coming from WSs freshly joining the federated
network,
without computing the b-metric.

\subsection{Performance evaluation\label{subsec:Bevaluation}}
We implemented the algorithm for evaluating the available bandwidth $B(j)$, the b-metric, as well as the automatic data rate
adaptation scheme AARF \cite{turletti} in the Omnet++~v4.1 simulator. 
To represent the propagation conditions over the wireless channel, 
we resort to a refinement of the ITU indoor
channel model, obtained using the experimental measurements presented in~\cite{ITU}. 
As for the algorithm parameters, 
we set  $T_{max}=0.1$~s. 

%

For clarity of presentation, here we consider only one IEEE 802.11g
BSS, including a Gateway and a varying number of WSs. All nodes can initially transmit at
54~Mbps and both elastic
(TCP) and inelastic (UDP) traffic flows are present. 
Also, since the available bandwidth $B(j)$ and the
b-metric are strongly linked to each other, we show the effectiveness of our approach in
predicting the first metric only.

Inelastic traffic is modeled as CBR flow with an offered load
of 8~Mbps.  We fix the payload size to 1500~bytes and, for clarity of
presentation we limit our study to 3 WSs.
Also, the depicted throughput is computed at the MAC layer and, for TCP traffic, it
 includes both data and TCP ACK packets. 

\begin{figure}[t]
\begin{center}
\subfigure[Aggregate throughput and $B(j)$ \label{fig:1TCPUDP_2UDP-tot}]
{\includegraphics [width=9cm] {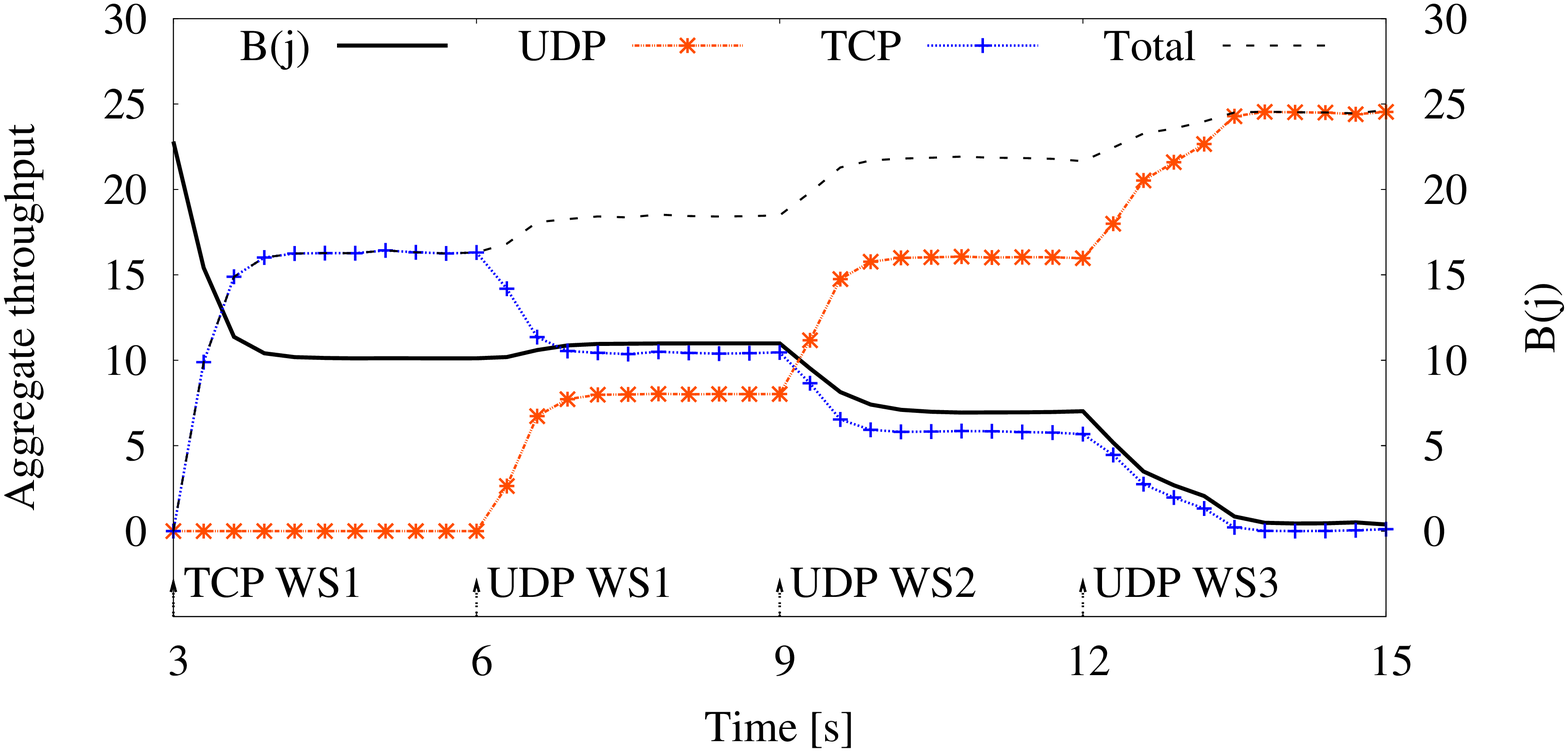}}
\subfigure[Per-WS throughput\label{fig:1TCPUDP_2UDP-single}]
{\includegraphics [width=8cm] {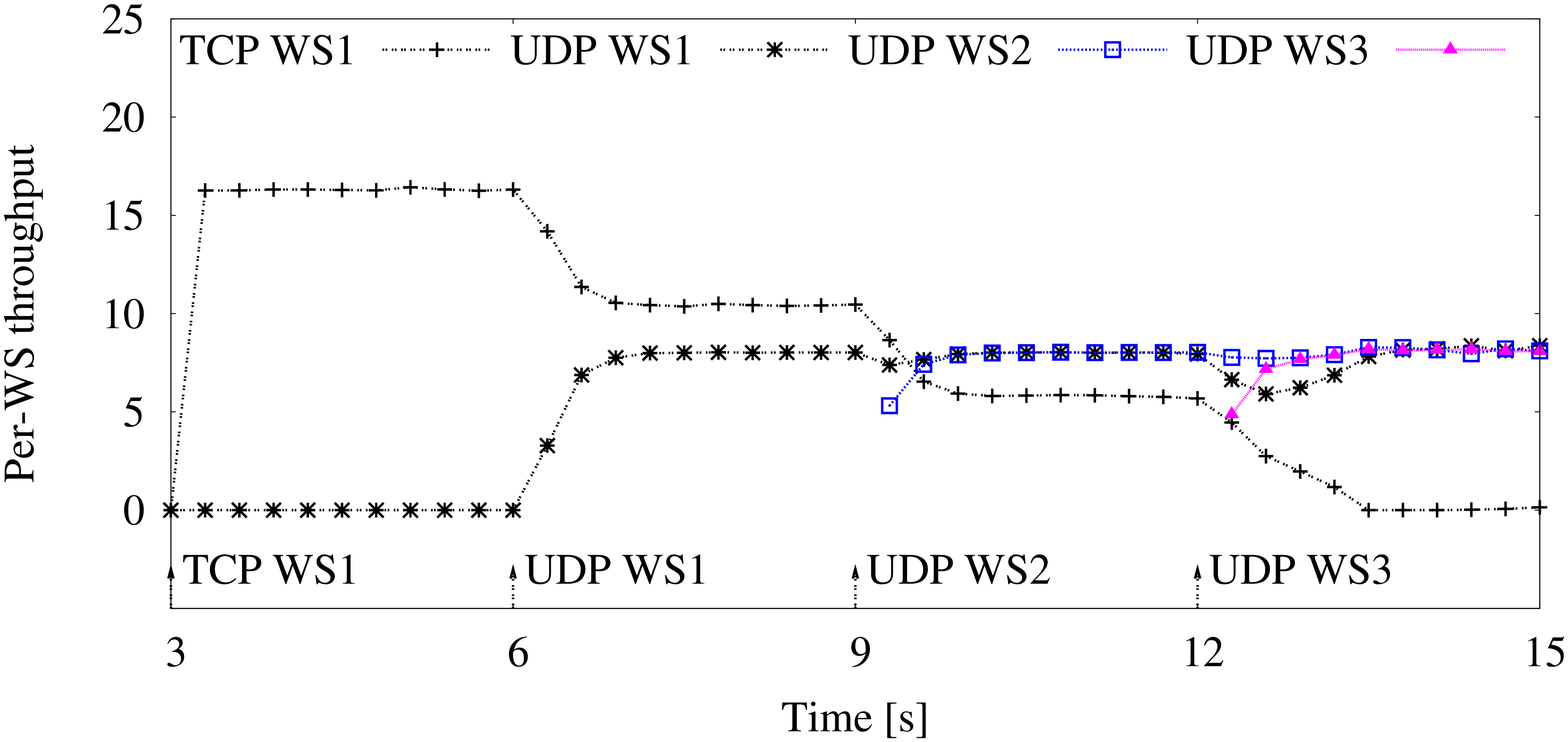}}
\caption{\label{fig:1TCPUDP_2UDP}WS~1 originates one TCP and one UDP
  flow, while WS~2 and WS~3 
originate one UDP stream each.
The flows become active at 3~s, 6~s, 9~s and 12~s, respectively.}
\end{center}
\vspace{-3mm}
\end{figure}
\begin{figure}[t]
\begin{center}
\subfigure[Aggregate throughput and $B(j)$ \label{fig:allTCPallUDP-tot}]
{\includegraphics [width=9cm] {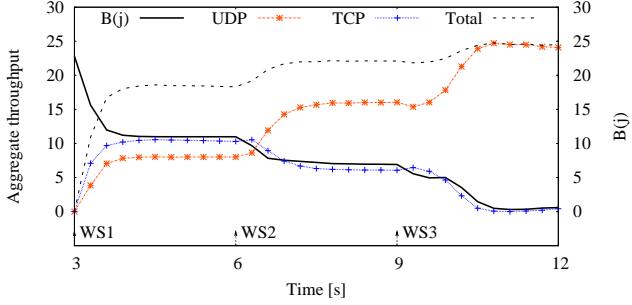}}
\subfigure[Per-WS throughput \label{fig:allTCPallUDP-single}]
{\includegraphics [width=8cm] {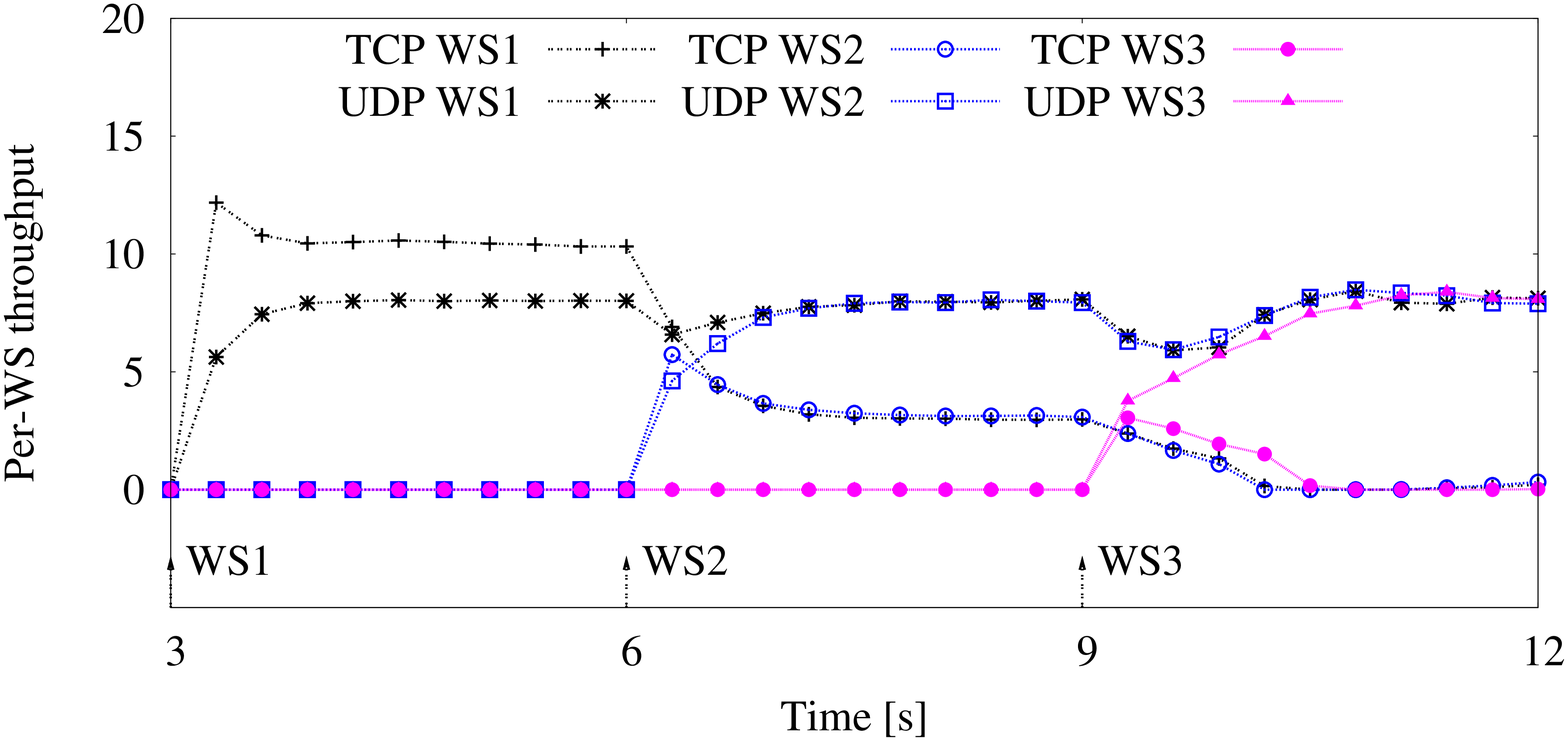}}
\caption{\label{fig:allTCPallUDP}Three WSs originate one TCP and one
  UDP flow each. The WSs become active at 3~s, 6~s and 9~s, respectively.}
\end{center}
\vspace{-3mm}
\end{figure}

We first consider that WS~1 starts a TCP
connection at $t=3$~s and, subsequently, a UDP flow at $t=6$~s. The
other two stations, WS~2 and WS~3, start a UDP flow at $t=9$~s 
and $t=12$~s,  respectively.
Fig.~\ref{fig:1TCPUDP_2UDP} shows the temporal evolution of the BSS
aggregate throughput and $B(j)$, as well as the throughput
achieved by  each WS. In spite of the saturation condition caused by
the 
TCP session started by WS~1 at $t=3$~s, $B(j)$ correctly
reflects that some bandwidth is available for the newly originated
flow. As the UDP stream starts at 6~s,  TCP adjusts its throughput and 
lets UDP take the desired bandwidth. Interestingly, we note that $B(j)$ is not 
significantly affected by this new condition. This is due to two
reasons: (i) the UDP stream is originated by the same WS that started
the TCP flow and (ii) the UDP demand is less than the estimated 
remaining bandwidth. The slight change that we observe in $B(j)$
results from the smaller number of TCP ACKs within the cycle, hence
from a greater observed average payload size. 
Conversely, when the UDP flow of WS~2 becomes active at $t=9$~s, $B(j)$ drops to 8~Mbps. 
The available bandwidth, though, is enough to accommodate the flow by
WS~3, which starts at $t=12$~s and brings the system to saturation, hence $B(j)$ drops
to 0. Also, as expected, 
the TCP flow almost dies out after  $t=12$~s.

We then assume that all WSs originate one UDP and one TCP flow each,
and that WS~1, WS~2 and 
WS~3 become active at $t=3$, 6 and 9~s, respectively. Due to the
competition between elastic and 
inelastic traffic within the same WS,  we expect that all TCP flows
will die out as the UDP 
streams accommodate their demand.  
Fig.~\ref{fig:allTCPallUDP} confirms such a guess showing that the time evolution
of the aggregate TCP throughput matches that of the bandwidth available for inelastic traffic; again, the $B(j)$ reflects such a behavior very closely. 

\begin{figure}[t]
\begin{center}
\subfigure[Aggregate throughput and $B(j)$ \label{fig:allTCP_dw_plusUDP_up-tot}]
{\includegraphics [width=9cm] {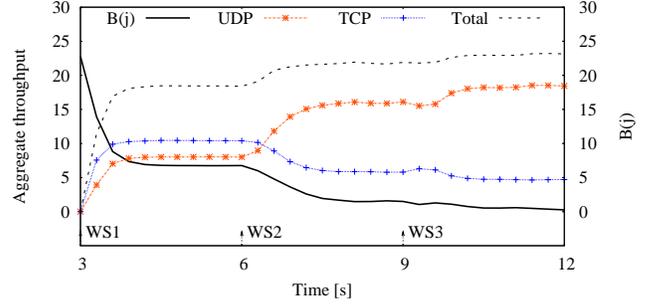}}
\subfigure[Per-WS throughput \label{fig:allTCP_dw_plusUDP_up-single}]
{\includegraphics [width=9cm] {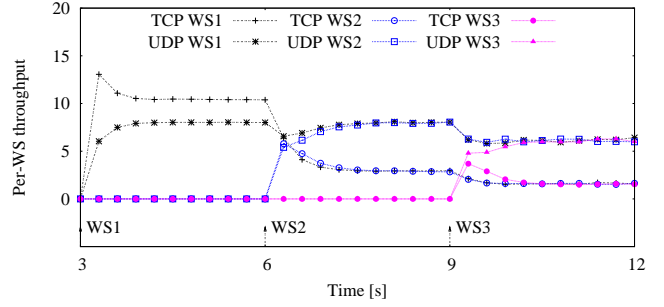}}
\caption{\label{fig:allTCP_dw_plusUDP_up}WS~1, WS~2 and WS~3 originate one UDP flow and are destinations of one TCP flow each. The WSs become active at 3~s, 6~s and 9~s, respectively.}
\end{center}
\vspace{-3mm}
\end{figure}

At last, we consider the same settings but for the TCP flows
direction: all WSs are now destinations of the TCP
traffic. Fig.~\ref{fig:allTCP_dw_plusUDP_up} shows that in this case
the UDP throughput equals the value of offered traffic only for $t \in
[3,6]$~s, i.e., when only WS 1 and the Gateway are active. In this time
interval, $B(j)$ correctly detects enough bandwidth to
accommodate an 8~Mbps-traffic flow. Then, by looking at
Fig.~\ref{fig:allTCP_dw_plusUDP_up-single}, we note that, after
$t=9$~s, both WS~1 and WS~2 suffer a loss with respect to their
demand, due to the new UDP flow started by WS~3. 
Consistently, $B(j)$ in Fig.~\ref{fig:allTCP_dw_plusUDP_up-tot}
indicates that no bandwidth was available for inelastic traffic. 
We point out that the throughput share of the Gateway, which is used for
TCP traffic, 
erodes some of the resources available for the WSs, 
due to the per-packet fairness provided by the DCF scheme. 


\section{Resource sharing protocol\label{sec:proto}}
In this section, we  describe our resource sharing protocol and 
show its performance in a residential scenario.

\subsection{Protocol description\label{subsec:proto}}
We now introduce the protocol that lets federated Gateways share
their radio resources. We remark that the presence of a central controller is
not required, and the implementation of the proposed protocol implies 
changes only at the Gateways, not in the WSs. 

\begin{figure}[!t]
\begin{center}
\includegraphics[width=8.5cm]{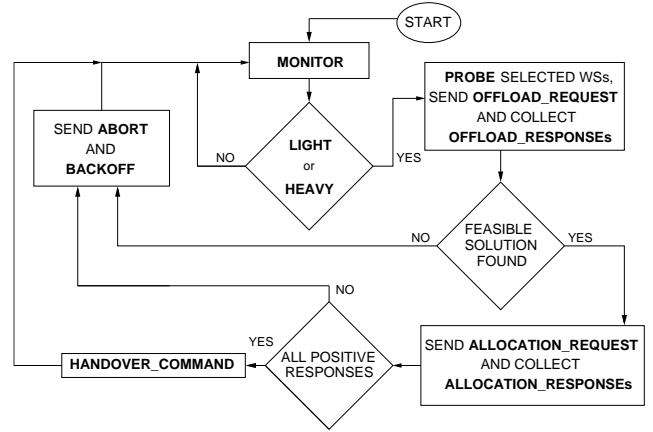}
\caption{Flow chart of the offload procedure.\label{fig:flow_chart}}
\end{center}
\vspace{-5mm}
\end{figure}

As already mentioned, our objective is twofold: (i) to minimize the
number of switched-on Gateways, and (ii) to avoid
overloading traffic conditions for the ``on'' Gateways.  
To achieve these goals, a Gateway periodically measures the load of
 its BSS  and assesses its status, as described
in Sec.~\ref{sec:algo}. 
If in Light or Heavy status, the Gateway carries out an offload
procedure, which is summarized in Fig.~\ref{fig:flow_chart}. 
The procedure aims at relocating one or more WSs at other 
Gateways. The federated Gateways 
estimate which WSs they could associate, based
on the value of their b-metric, and reply 
accordingly. Upon finding
a valid WS relocation, the Gateway that started the procedure can turn itself 
off if it was in Light status, while it experiences a load decrease if
it was in Heavy status.
The procedure for a Gateway in
Light or Heavy status is detailed below.

{\bf Light status.} Consider a Gateway $G_l$ that finds itself in Light status. Then, $G_l$
starts an offload procedure by multicasting 
an {\sc offload\_request} message to the federated
Gateways. This message includes the status of the requesting Gateway, 
the frequency channel currently used in the BSS, 
a hash of the association ID (AID), 
the MAC address and the measured traffic profile of each WS in the BSS.
After the {\sc offload\_request} is issued, $G_l$  sets a timer to the
timeout value $\tau_r$.

An {\sc offload\_request} is processed only by federated Gateways that
are currently  on and not in Heavy status.
Since the request comes from a Gateway in Light status, 
the federated Gateways first check if their b-metric is greater than the value
advertized by $G_l$. If so, they discard the request since they are
less loaded than $G_l$. 
Otherwise, they need to evaluate which of the WSs are in their
radio range and which data rate they could use to communicate with
the WSs. To do so, we let the Gateways tune one of their radio
interfaces 
to the channel used by $G_l$ for a time $\tau_p$; then, we let 
$G_l$ probe each WS in its BSS with an RTS message. 
As the probed WS will reply with a CTS, the Gateways monitoring the
frequency channel will be able to estimate the signal-to-noise ratio
(SNR), hence the data rate they could use to communicate with the WS.
Note that $G_l$ will set the RTS duration field so that the corresponding field
in the CTS will be the hash function of the WS's AID\footnote{The RTS
  duration field is set to the sum of the SIFS time, CTS
  transmission time and the hash of the WS's AID. The value of the
  hash should be upper bounded by  $2 \cdot \mbox{SIFS}$  plus the ACK
duration so that probe CTS cannot be mistaken with regular CTS.}. 
Such a procedure allows a Gateway that is not in radio proximity of
$G_l$ (i.e., unable to hear the RTS) to identify the WS sending the
CTS. Clearly, it introduces some overhead, but, since
$G_l$ is underloaded, we expect the number of  WSs in its BSS to be small. 
 
Each federated Gateway then considers the WSs from which is has heard
a CTS within the time $\tau_p$. To verify which WSs (if any) could be associated to
its BSS, 
the Gateway evaluates through  Alg.~\ref{algo:ACE} the b-metric for the possible combinations 
of candidate WSs. 
Finally, it unicasts an {\sc offload\_response} message to $G_l$, including 
the combinations with a positive outcome (i.e., $\mbox{b}(j)>0$), 
as well as the corresponding value of the b-metric and the data rates that
could be used to communicate with the candidate WSs.

Upon the expiration of the timeout $\tau_r$, $G_l$ evaluates all received
replies. 
Among the feasible solutions, the allocation  maximizing the
average data rate of the WSs is selected. 
To solve possible ties, preference is given to the allocation that minimizes the average
b-metric. 
The rational is that, firstly, WSs should be handed over to the Gateways that
will be able to communicate with them at the highest data rate, so as
to guarantee an efficient traffic transfer. Secondly, we want 
as many WSs as possible to
associate to Gateways that have already a high
traffic load and leave out those that are likely to reach a Light
status, hence to switch themselves off. 

If a valid allocation is found, $G_l$ unicasts to 
each selected Gateway  an {\sc  allocation\_request}, including the MAC address of the WSs assigned
to it  and the current b-metric value of $G_l$. 
A Gateway  receiving the {\sc
  allocation\_request}  evaluates again the b-metric taking the
assigned WSs into account. 
If the result of the evaluation is still positive and its b-metric is
less than the value advertized by $G_l$, the Gateway replies with a positive {\sc
  allocation\_response}; otherwise, it sends a negative {\sc
  allocation\_response}.
$G_l$ will end the offload procedure by multicasting to all Gateways a {\sc
  handover\_command} if it receives all positive {\sc
  allocation\_response}s, or an {\sc abort} message otherwise. 
Upon the reception of a {\sc  handover\_command}, each selected
Gateway will include the assigned WS(s) in its authorized stations
list, so that, when $G_l$  switches itself off, each WS will necessarily
associate with the right Gateway. 

{\bf Heavy status.} When a Gateway, $G_h$, finds itself in 
Heavy status, it  starts an offload procedure similar to the one
described above. 
A few differences, however, exist. Firstly, $G_h$ tries
to hand over only one WS at a time, till its status changes into Regular.
Specifically, it lists the WSs in decreasing order
according to their offered load weighted by the inverse of their data
rate, and attempt to relocate the top WS
first. Thus, the handover of each WS results in a different
offload procedure. Secondly, upon receiving an {\sc offload\_request}
from $G_h$, an ``on''
Gateway not in Heavy status will always reply, provided that its b-metric computed through
Alg.~\ref{algo:ACE} is greater than 0. 
However, if no viable relocation is found, $G_h$ will resend 
the {\sc offload\_request} with a flag set. This message will be
processed also by 
``off'' Gateways, with a given
probability 
(low-power circuits \cite{lowpow-wakeup} can be  used to wake up
Gateways  upon the reception of the message with the flag set). 
In this way, we let ``off'' Gateways turn themselves on if needed, while
limiting the number of Gateways that wake up.

We remark that, 
upon receiving an {\sc offload\_request}, a  Gateway wishing to start an
offload procedure  defers its request till it receives a 
{\sc handover\_command}  or an {\sc abort}, and then backoff.
This ensures that in the federated network there is only one active offload procedure at the time.


\subsection{ Performance evaluation} 
\label{sec:Pevaluation}
\begin{figure}[!t]
\begin{center}
\includegraphics[width=6cm]{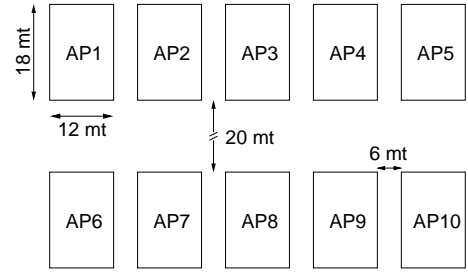}
\caption{Federated detached houses scenario.\label{fig:scenario}}
\end{center}
\vspace{-5mm}
\end{figure}

We implemented our protocol in the Omnet++~v4.1 simulator and 
evaluated its performance in a realistic scenario referring to 
a neighborhood located in the suburbs
of Chicago, IL. 
The scenario, depicted in  Figure~\ref{fig:scenario}, includes 10
federated detached houses, 
each equipped with an IEEE 802.11g Gateway.
As previously mentioned, channel propagation conditions are modeled through 
the model defined in~\cite{ITU}. 
Also, the average fraction of Gateways in
radio visibility of a WS, when a data rate of 1~Mbps is used, is 
0.8.  
As for the algorithm parameters, we have $T_R=0.05$, $T_L=0.5$, $T_A=0.2$, $N_L=10$,
$T_{max}=0.1$~s, $\tau_r=0.3$~s, and $\tau_p=0.1$~s, while we set to 0.5 the
probability
that an ``off'' Gateway turns itself on upon receiving 
a flagged {\sc offload\_request} from a neighboring Gateway in Heavy
status.

For reasons of space, we limit the set of results to a scenario featuring
only uplink UDP traffic.  
Consequently, we set the
offload procedure to be quite reactive (a few seconds in Light/Heavy status are sufficient
to trigger it). 
Additional hysteresis (i.e., heavier smoothing when 
computing running averages of throughput) is needed to cope with the 
periodic fluctuations of TCP flows.

\begin{figure}[t]
\begin{center}
\subfigure[Gateways throughput (Light status)\label{fig:Throughput_din_1}]
{\includegraphics [width=8.5cm] {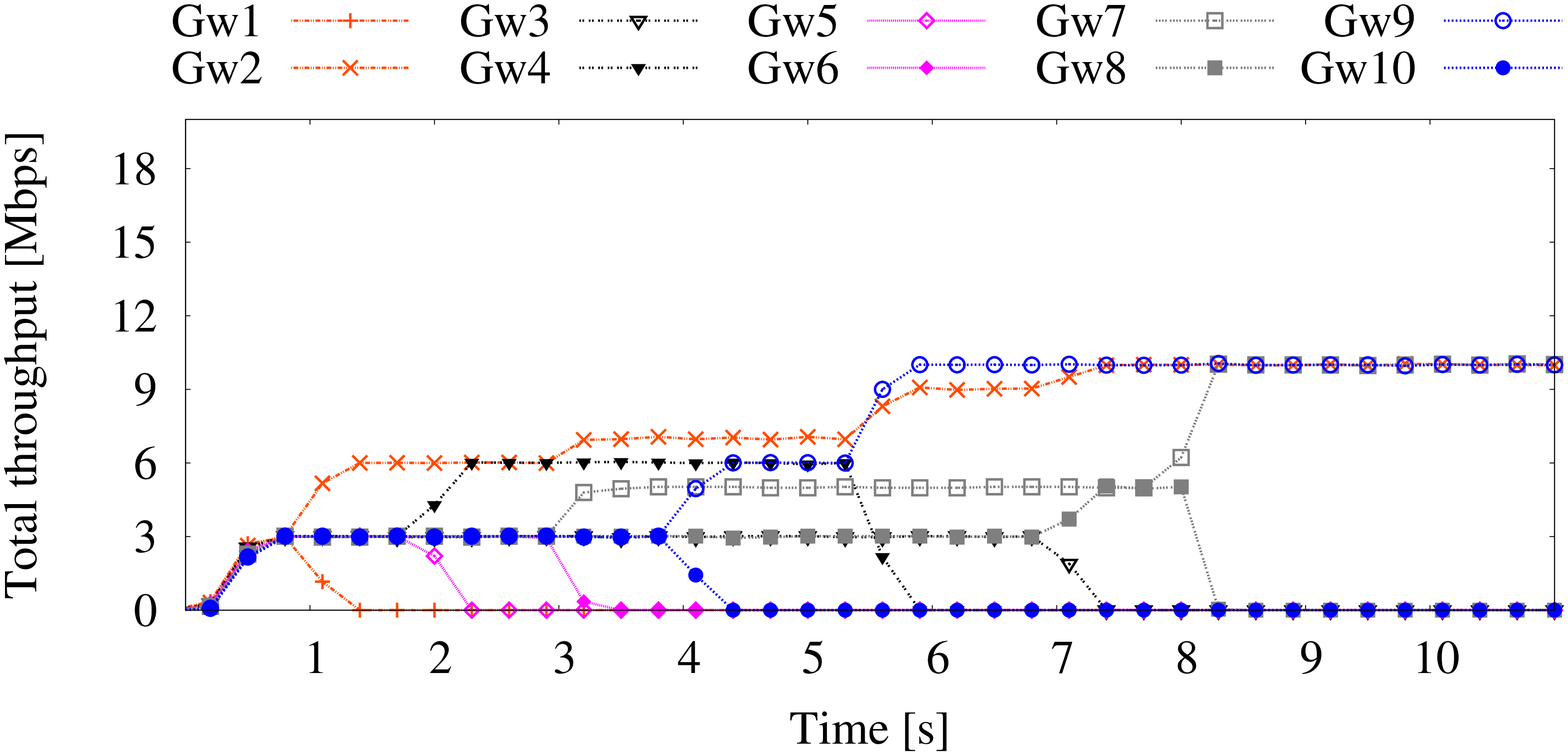}}

\subfigure[Gateways throughput (Heavy status)\label{fig:Throughput_din_2}]
{\includegraphics [width=8.5cm] {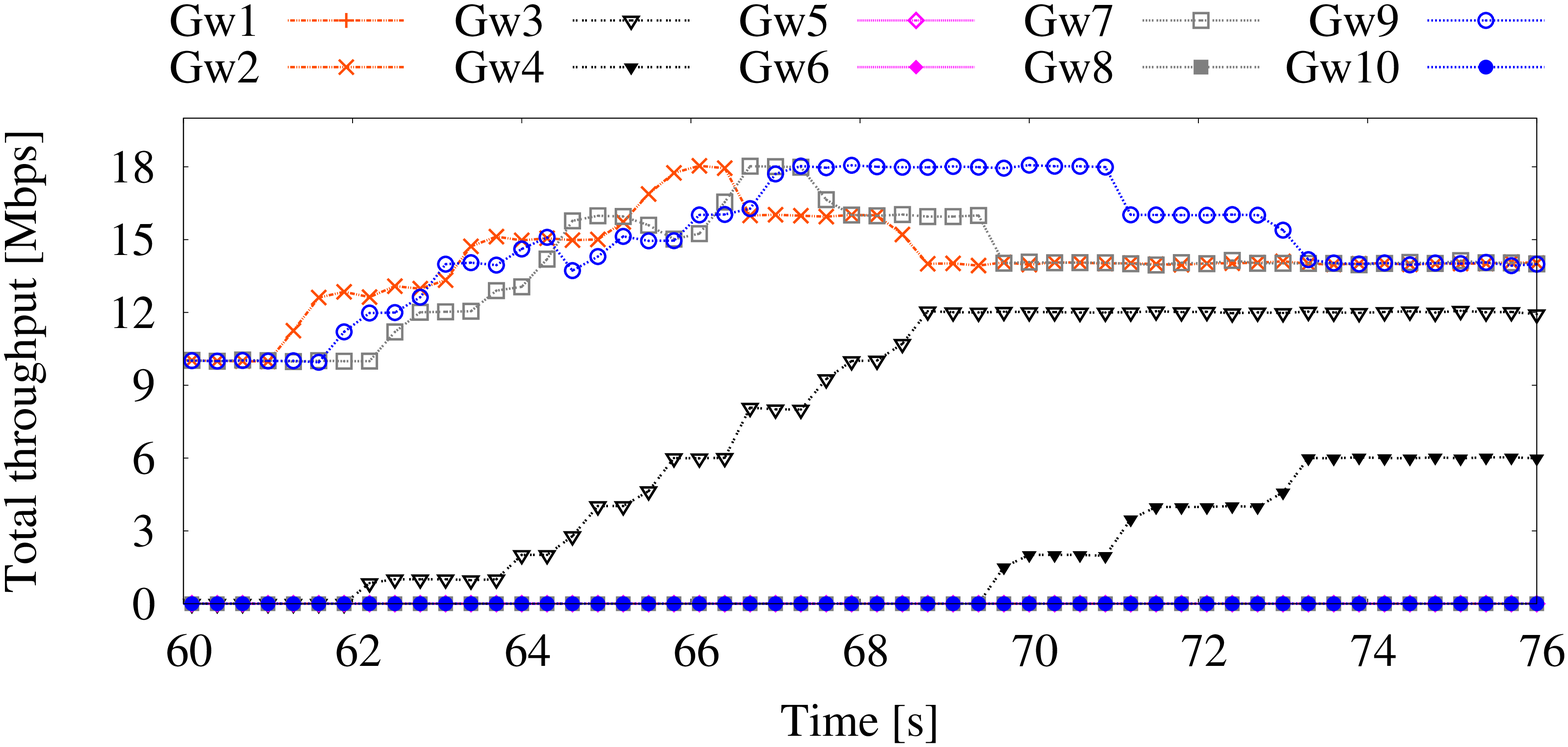}}

\caption{\label{fig:dynamic}Temporal evolution of the Gateways
  throughput under Light and Heavy conditions.}
\end{center}
\vspace{-5mm}
\end{figure}

In order to evaluate the behavior of our scheme in  Light and
Heavy status, we consider a dynamic traffic scenario. Initially, 
all Gateways are ``on'' and they have 3  associated WSs each. 
At time $t$=0~s, every WS starts generating an uplink UDP stream at
1~Mbps (see Fig.~\ref{fig:Throughput_din_1}); since the  per-Gateway load is 3~Mbps, all Gateways are in Light
status. Then, between 60 and 68~s, every WS doubles its offered load
(see Fig.~\ref{fig:Throughput_din_2}), driving the ``on'' Gateways into  Heavy status. 

\begin{figure}[!t]
\begin{center}
\includegraphics[width=8.5cm]{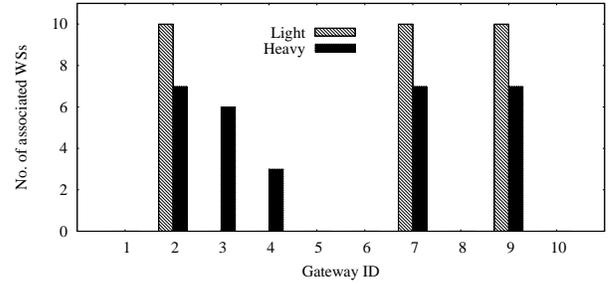}
\caption{WS distribution over the Gateways under Light and Heavy conditions.\label{fig:WS_distr}}
\end{center}
\vspace{-5mm}
\end{figure}

The temporal evolution of the Gateways  throughput, when  all
Gateways are initially in Light status,  is
shown in Fig.~\ref{fig:Throughput_din_1}, where different 
marker/color  combinations are
used to represent the behavior of single Gateways. The Gateways
that successfully carry out an offload procedure and become  
``off'' correspond to downward curves, while Gateways that associate relocated
WSs see their throughput grow. A sample of a successful offload can be
observed in the interval $[3,4]$~s 
where a Gateway, upon switching itself off, relocates its three WSs to two other
Gateways whose throughput therefore increases. 
Eventually (at $t$=8.5~s), the federated network settles at 3 ``on'' 
Gateways out of 10. Each ``on'' Gateway
serves 10 WSs (see Fig.~\ref{fig:WS_distr}) and is in Regular
status.

Then, Fig.~\ref{fig:Throughput_din_2} shows the temporal evolution
of the Gateways  throughput when a sudden traffic increase drives the
three 
``on'' Gateways into Heavy status.
As the WSs progressively double their offered load (between 60 and
68~s), two additional  Gateways turn themselves on and come to the aid
of the overloaded ones. We remark that the proposed algorithm always tries to 
minimize the number of ``on'' Gateways, thus the second one is
switched on only when the first  can no longer associate WSs without
moving into Heavy status itself.
 When all Gateways are in  Regular status
($t$=73~s), no further relocations occur and the network
stabilizes at 5 ``on'' Gateways.
The three Gateways that were ``on'' at the end of the period depicted in 
Fig.~\ref{fig:Throughput_din_1} now have  7 associated WSs, while the
first and the second Gateway that came in aid accepted 6 and 3 WSs,
respectively, as shown in Fig.~\ref{fig:WS_distr}.

%


\begin{figure}[t]
\begin{center}
\subfigure[Percentage of ``off'' Gateways\label{fig:Villette_numWS}]
{\includegraphics [width=8.5cm] {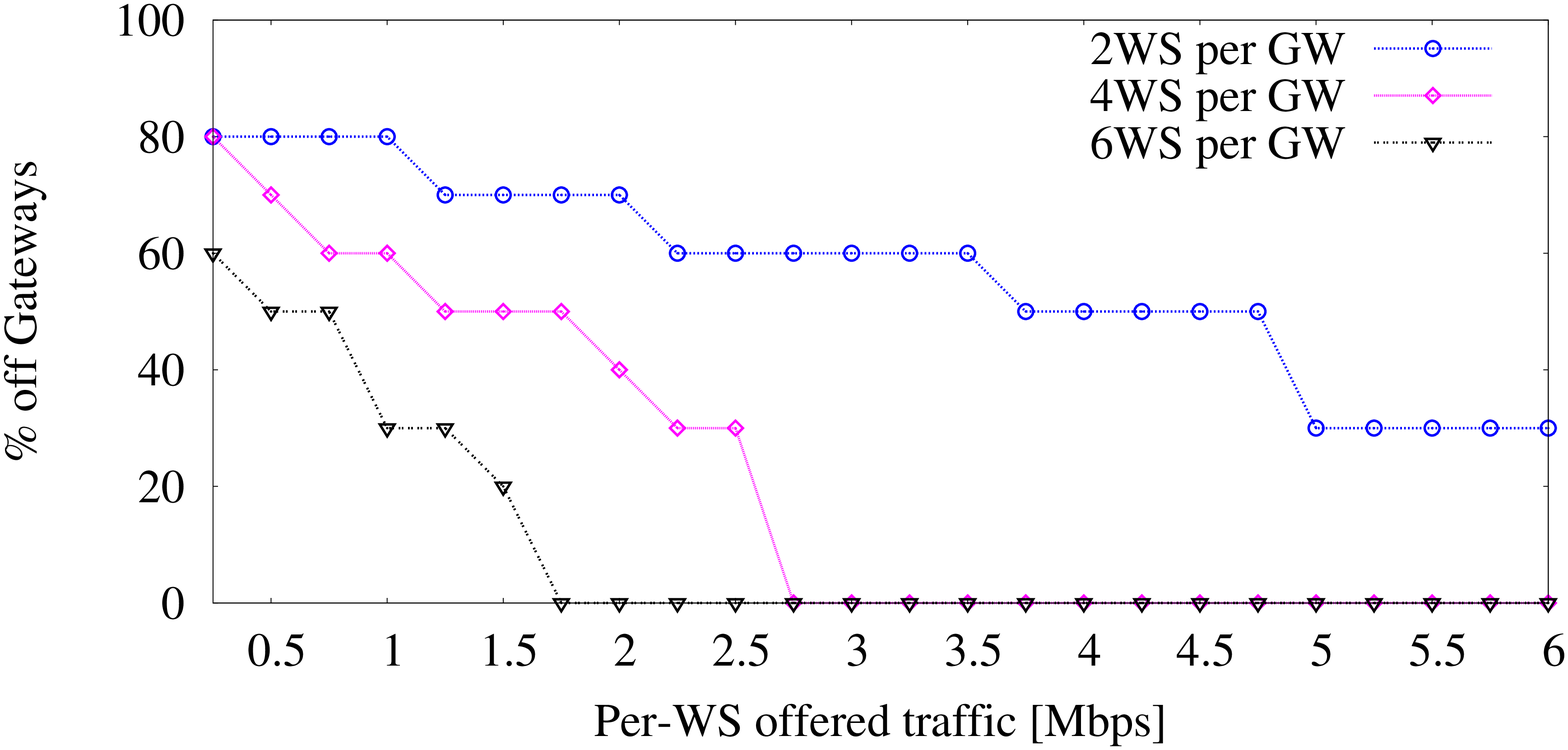}}
\subfigure[Average number of WSs/Gateway\label{fig:Villette_numWS_avgmax}]
{\includegraphics [width=8.5cm] {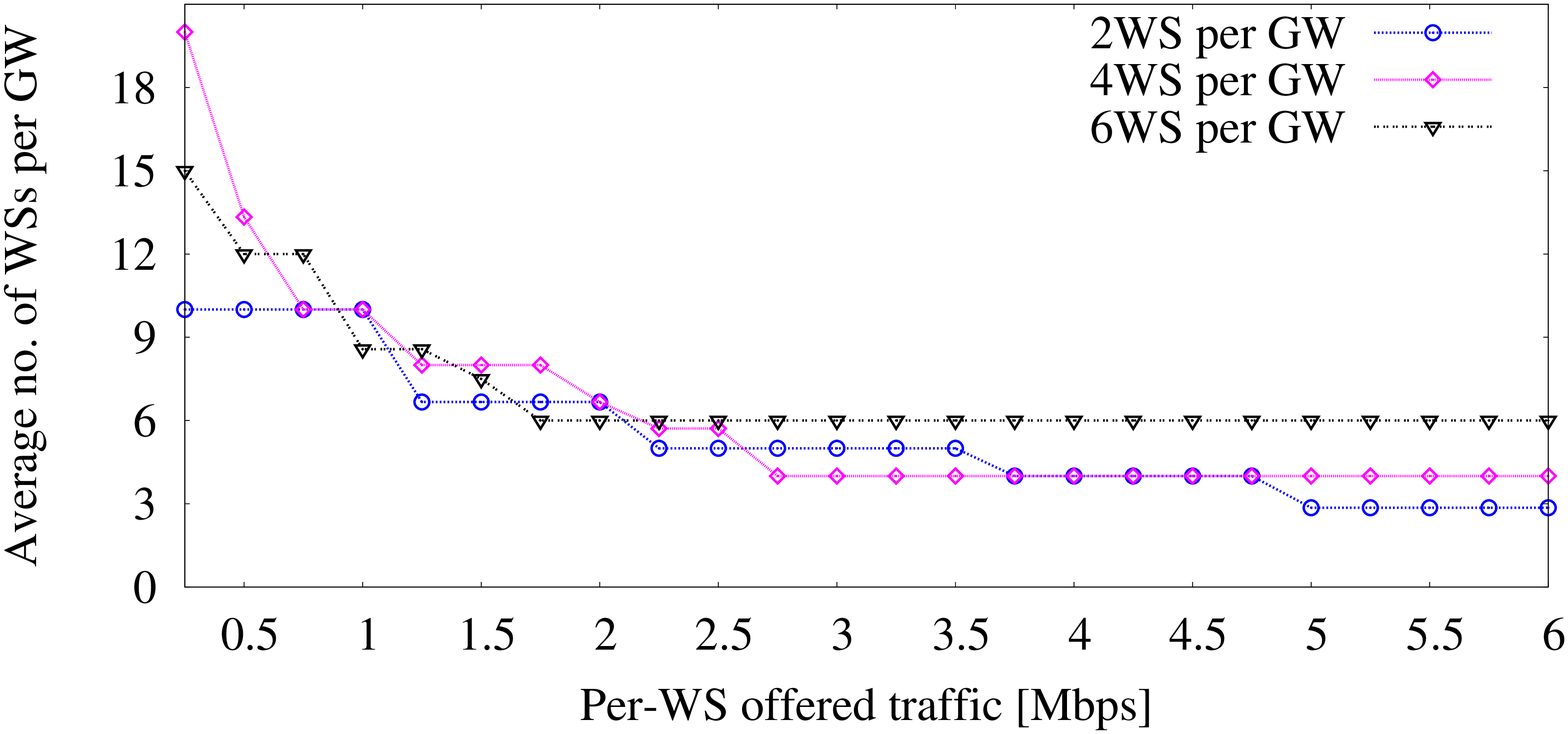}}
\caption{\label{fig:swipe} Percentage of ``off'' Gateways and average
  number of associated WSs per Gateway, as the WS offered load
  varies and for a different initial number of WSs per Gateway.}
\end{center}
\vspace{-5mm}
\end{figure}

Next, we consider a different traffic scenario where initially all 10
Gateways serve the same number of WSs (namely, 2, 4, 6). Each WS 
generates a UDP flow with the same offered load, which is a varying
parameter in different test runs. 
Fig.~\ref{fig:swipe} shows the percentage of ``off'' Gateways, as well
as the average number of WSs associated to a Gateway, upon reaching
steady state.
As expected, the number of switched off Gateways decreases as both
the offered load and the number of WSs in the
federated network increase. These results suggest that, for widely
different load conditions, the configuration yielded by our solution
well adapts to the system dynamics.

\section{Conclusion and future work} 
\label{sec:future}

We designed a set of procedures aimed at managing underload and overload
conditions in wireless Gateways of federated households. After outlining some 
methodologies for throughput monitoring in presence of uplink/downlink elastic and inelastic
traffic, we introduced the offload procedures that allow (i) an underloaded Gateway to 
relocate all of its WSs and thus switch off; (ii) an overloaded Gateway to relocate 
some of its WSs and alleviate its status. By simulation, we then showed the effectiveness
of the procedures in a simple, yet realistic federated neighborhood scenario.

Further developments will address a wider evaluation of federated scenarios in presence of TCP traffic, prompt management of ``off'' Gateways, 
as well as power saving benchmarks comparing our solution with an always-on
Gateway setting. The implementation of our solution in real devices
will follow, along with experimental measurements.




\section*{Appendix}  
The average time duration of a possible event taking place on the channel is given by:
\begin{equation}
\begin{split}
E[T(j)] = (1-\tau(j))^{N(j)}~\sigma + \\
[N(j) \tau(j)(1-\tau(j))^{N(j)-1}(1-p_e(j))] T_s(j) + \\
[1-(1-\tau(j))^{N(j)}-N(j)\tau(j)(1-\tau(j))^{N(j)-1}] T_c(j) + \\
[N(j) \tau(j)(1-\tau(j))^{N(j)-1}p_e(j)] T_e(j)
\label{eq:T}
\end{split}
\end{equation}
where $\sigma$ is the slot time duration.
The average duration of a successful transmission, $T_s(j)$, and of 
an erroneous transmission, $T_e(j)$, are derived as follows: 
\begin{eqnarray}
 \hspace{-3mm} T_s(j)  \hspace{-3mm} &=&  \hspace{-3mm} 2\frac{h_{phy}}{R_b}  +\frac{h_{mac}+P(j)+\mbox{\sc ack}}{R(j)} +  \mbox{{\sc sifs}} +  \mbox{\sc difs}
\label{eq:Ts} \\
 \hspace{-3mm} T_e(j) \hspace{-3mm} &=&  \hspace{-3mm} \frac{h_{phy}}{R_b}+\frac{h_{mac}+P(j)}{R(j)} + T_o + \mbox{\sc difs} \,.
\label{eq:Te}
\end{eqnarray}
In (\ref{eq:Te}), $h_{phy}$ is the length of the physical header for the data and
the ACK frames (assumed to be transmitted at the basic rate $R_b$),
$h_{mac}$ is the length of the MAC header, {\sc ack} is the length of the
acknowledgment MAC fields and $T_o$ is the retransmission timeout,
which we set equal to  {\sc sifs} plus the ACK duration.
As for the exact computation of the average collision duration, the Gateway should be aware of the number of nodes that are hidden with respect to each other. The works in~\cite{bianchi,chatz} do not account for hidden WSs and the approaches proposed in the literature
are not viable in our set up, as we do not require the Gateway to have knowledge of the users distribution within its coverage area.
Thus, we approximate the average collision duration by making the following worst-case assumption: each collision in cycle $j$ involves a packet of maximum size $P_{max}(j)$; then
\begin{equation}
T_c(j)=\frac{h_{phy}}{R_b}+\frac{h_{mac}+ P_{max}(j)}{R(j)} + T_o + \mbox{{\sc difs}} \,.
\end{equation}
Clearly, the above expression  may lead to overestimating the average collision time in absence of hidden terminals, hence to underestimating the theoretical saturation throughput; this, however, is acceptable for our purposes, as also proved by the simulation results presented 
in Sec.~\ref{subsec:Bevaluation}. 

We also observe that the Gateway can easily compute $\tau(j)$ using the following equation\cite{chatz}:
\begin{eqnarray}
p(j)&=&1-[(1-\tau(j))^{N(j)-1}(1-p_e(j))] \\
\tau(j)& = &1 + \bigg[\frac{p(j)-1}{1-p_e(j)}\bigg]^{\frac{1}{(N(j)-1)}}
\end{eqnarray}
where $p(j)$ is the the conditional probability that a transmitted packet encounters a collision or is received in error in saturation conditions.
Note that $p(j)$ and $\tau(j)$ have to be obtained through numerical methods, as described in
\cite{chatz,bianchi}.

\end{document}